\documentclass[sigconf,authorversion]{acmart}
\usepackage{makecell}
\usepackage{ragged2e}
\usepackage{longtable}
\usepackage[normalem]{ulem}
\usepackage{graphicx}
\usepackage{booktabs}
\usepackage{threeparttablex}
\usepackage[dvipsnames]{xcolor}
\usepackage{amsmath}
\usepackage{pifont}

\usepackage[width=.99\textwidth]{caption}
\usepackage{rotating}
\usepackage{pdflscape}
\usepackage{float}
\usepackage{multirow}
\usepackage{amsmath}
\usepackage{rotating}
\usepackage{tikz}
\usepackage{dblfloatfix}
\usepackage{caption}
\usepackage{subcaption}
\usepackage[ruled]{algorithm2e}
\usepackage{listings}
\usepackage{xcolor,pifont}
\usepackage{booktabs}
\usepackage{multirow}
\AtBeginDocument{%
  }

\setcopyright{acmlicensed}
\copyrightyear{2026}
\acmYear{2026}
\acmDOI{XXXXXXX.XXXXXXX}
\acmConference[ICFNDS '26]{}{Nov 04--06,2026}{Izmir, Turkey}
\acmISBN{978-1-4503-XXXX-X/2018/06}

\usepackage[nointegrals]{wasysym}

\newcommand{\level}[1]{%
\ifcase#1
\or \Circle      
\or \LEFTcircle  
\or \CIRCLE      
\fi
}

\usetikzlibrary{positioning, shapes}

\tikzset{
  simnode/.style={
    draw, rectangle, rounded corners=2pt,
    minimum width=0.7cm, minimum height=0.45cm,
    fill=blue!15, font=\scriptsize
  },
  emunode/.style={
    draw, rectangle, rounded corners=2pt,
    minimum width=0.7cm, minimum height=0.45cm,
    fill=orange!20, font=\scriptsize
  },
  simlink/.style={draw, thick, blue!60},
  bridgelink/.style={draw, thick, dashed, purple!70},
  emulink/.style={draw, thick, orange!80!black},
  linklabel/.style={font=\tiny, inner sep=1pt},
}

\definecolor{simblue}{RGB}{30,100,200}
\definecolor{emugreen}{RGB}{30,160,80}
\definecolor{bridgeorange}{RGB}{220,120,20}
\definecolor{codegray}{RGB}{245,245,245}
\definecolor{codeblue}{RGB}{0,0,180}
\def \ProjectName {AgenticNet}

\usepackage[most]{tcolorbox}
\newtcolorbox{gptprompt}[1]{
  enhanced,
  breakable,
  colback=blue!5,           
  colframe=blue!70,        
  colbacktitle=blue!80,     
  coltitle=white,         
  fonttitle=\bfseries\small,
  title={#1},
  boxrule=0.8pt,
  arc=2mm,
  left=2mm,
  right=2mm,
  top=1mm,
  bottom=1mm,
  boxsep=1mm,
  width=\columnwidth
}

\definecolor{maroon}{RGB}{220,0,0}

\newtcolorbox{gptprompt2}[1]{
  enhanced,
  breakable,
  colback=red!5,
  colframe=maroon,
  colbacktitle=maroon,
  coltitle=white,
  fonttitle=\bfseries\small,
  title={#1},
  boxrule=0.8pt,
  arc=2mm,
  left=2mm,
  right=2mm,
  top=1mm,
  bottom=1mm,
  boxsep=1mm,
  width=\columnwidth
}

\lstdefinestyle{switchstyle}{
    backgroundcolor=\color{blue!10},    
    basicstyle=\ttfamily\footnotesize,
    keywordstyle=\color{purple}\bfseries,    
    classoffset=1,   
    morekeywords={MySwitch,ForwardingModule,process,on_attach,reset},
    keywordstyle=[1]\color{blue!70}\bfseries,  
    commentstyle=\color{green!60!black}\itshape, 
    stringstyle=\color{red!80!black},       
    frame=single,
    rulecolor=\color{blue!50},
    breaklines=true,
    showstringspaces=false,
    numbers=none,
    xleftmargin=2mm,
    xrightmargin=2mm
}

\begin{document}


\title{\ProjectName: Utilizing AI Coding Agents To Create Hybrid Network Experiments}
\author{Majd Latah}
\affiliation{%
  \institution{Ozyegin University}
  \city{Istanbul}
  \country{Turkey}}
\email{majd.latah@ozu.edu.tr}

\author{Kubra Kalkan}
\affiliation{%
  \institution{Ozyegin University}
  \city{Istanbul}
  \country{Turkey}}
\email{kubra.kalkan@ozyegin.edu.tr}

\renewcommand{\shortauthors}{Majd Latah and Kubra Kalkan}

\begin{abstract}
Traditional network experiments focus on validation through either simulation or emulation. Each approach has its own advantages and limitations. In this work, we present a new tool for next-generation network experiments created through Artificial Intelligence (AI) coding agents. This tool facilitates hybrid network experimentation through simulation and emulation capabilities. The tool supports three main operation modes: pure simulation, pure emulation, and hybrid mode. \ProjectName{} provides a more flexible approach to creating experiments for cases that may require a combination of simulation and emulation. In addition, \ProjectName{} supports rapid development through AI agents. We experimentally evaluate the tool and present an approach to verify the generated code.
\end{abstract}

\ccsdesc[500]{Networks~Network simulations}
\ccsdesc[500]{Networks~Network experimentation}


\maketitle

\section{Introduction}
Existing network experiment tools focus either on simulation or emulation. For instance, Mininet\footnote{Mininet, \url{https://github.com/mininet/mininet}} provides emulation for SDN-focused experiments. OMNET++\footnote{OMNET++, \url{https://github.com/mininet/mininet}} is useful for simulations. There is a need for tools that combine both simulation and emulation. Those tools are essential for the research community. In this paper, we leverage AI coding agents to provide a proof-of-concept for hybrid network experiments. This AI-based simulator, AgentNet, provides the following benefits:

\begin{itemize}
  \item \textbf{Hybrid Experiments:} built-in support and coexistence of simulated and emulated nodes within the same network experiment.
  \item \textbf{Fine-grained Control:} per-link parameters (delay, bandwidth, loss), per-node failure, traffic isolation, and custom code.
  \item \textbf{Rapid development through Large Language Model (LLM)-based agents:} an interactive LLM-based agent to provide the ability to rapidly create network experiments using natural-language descriptions.   
\end{itemize}

Existing tools focus either on simulation or emulation.
This includes network simulators such as OMNET++ and ns-3\footnote{Ns3, \url{https://www.nsnam.org/}}. In addition to emulators such as Mininet.  In our work, we consider AI-based approach to create a hybrid network simulator, where both simulation and emulation nodes can be added through a simple API call. We also consider the ability to generate and verify the experiment code through AI agents. Our contributions are as follows:

\begin{itemize}
	\item We explore the utilization of AI-coding agents to create a hybrid network simulator.
	\item We evaluate the tool through network experiments.
	\item We also introduce an agent-assisted approach to verify the code generated for a given network experiment.
\end{itemize}

\section{Related-work}

\cite{fall1999network} provides early work on adding emulation capabilities to the NS simulator. To support emulation in the NS simulator, the scheduler is modified to allow real-time synchronization. Furthermore, packet capture and generation capabilities are needed to connect the simulation to the real-world. In \cite{guruprasad2005integrated}, the authors combine simulated nodes using \cite{fall1999network} in the Emulab testbed. In \cite{manishankar2019technologies}, the authors combine OMNET++ simulation with virtual nodes created through the Xen hypervisor. The authors suggest that the simulation model can use the UDP protocol to interact with the virtual nodes. \cite{jin2012virtual} introduces an integration of OpenVZ-based emulation into an S3F-based \cite{nicol2011s3f} simulator. The authors introduce global synchronization and a control mechanism, called the virtual environment (VE) controller. A global scheduler is added to coordinate the time advancement of emulated and simulated nodes. In \cite{hannon2018combining}, the author combined a power distribution system simulator (OpenDSS) and Mininet for smart grid environments. A virtual time system is added for synchronization between emulated and simulated parts.
In \cite{jin2015parallel-sdn}, the authors focus on integration between simulation and emulation of SDNs. The authors provide a global synchronization algorithm based on virtual time \cite{jin2012virtual}.
In another work \cite{liu2015toward},
the authors improve the capabilities of Mininet 
through a symbiotic approach, which combines emulation and simulation for enabling large-scale hybrid experiments. A traffic monitor and a traffic generator are added to the simulator to synchronize with Mininet instances. Furthermore, the communication between Mininet and the simulator is done using the TCP protocol.


\section{\ProjectName{}}
Initially, we asked Claude Code to design a simulator that combines the benefits of simulator and emulator frameworks (See Prompt 1).

\begin{gptprompt}{Prompt 1: Creating \ProjectName}
\textit{We need to design a simulator that combines the flexibility of ns-3 or OMNET++ (i.e., simulation) and the realistic approach of Mininet (i.e., emulation).}
\vspace{1mm}
\end{gptprompt}

\begin{figure}[h!]
	\centering
	\includegraphics[scale=0.31]{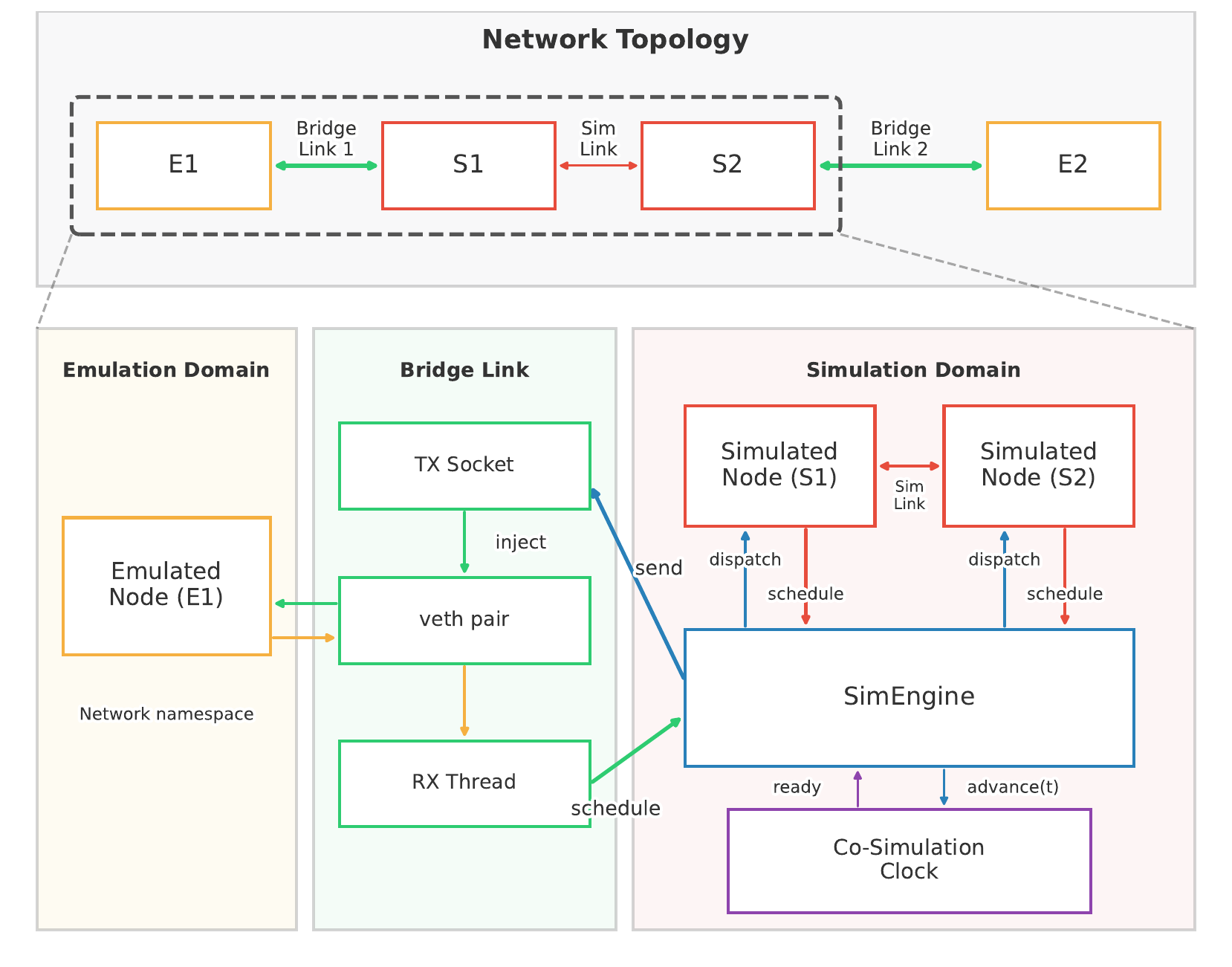}
	\caption{\ProjectName{} (designed using Claude Code)}
	\label{SDNLayer}
\end{figure}

First, we explore the main design suggested by the coding agent (see Fig. \ref{SDNLayer}). This design consists of four main components:

\begin{itemize}
  \item \textbf{Simulation Domain:} pure simulation.
  \item \textbf{Emulation Domain:} pure emulation.
  \item \textbf{Bridge Link:} connects the simulation domain to the emulation domain.
  \item \textbf{Time Synchronization:} the co-simulation clock synchronizes virtual simulation time with wall-clock time, which ensures time-consistent interaction between the simulation and emulation domains.
\end{itemize}

The tool supports the following three main working modes as follows. \textbf{Pure simulation mode} supports only simulations. \textbf{Pure emulation mode} supports only emulations. \textbf{Hybrid mode} supports both simulation and emulation.

\textbf{Emulation-to-simulation direction}:
the packets arriving from the emulation domain are timestamped with the virtual time of the simulation engine. Those packets are placed in a shared event queue and then dispatched once wall-clock time reaches the corresponding timestamp within a bounded synchronization window. 

\textbf{Simulation-to-emulation direction}:
no additional wait is needed for packets leaving the simulation domain, since the emulation domain is synchronized with the wall-clock.

\section{Testing \ProjectName}

\subsection{Exp1: Comparing different working modes}
We evaluate the tool across three operational modes using the same topology (see Fig. \ref{Expl1-Topology}): a source node transmits packets through a switch to a destination, with a 5 ms link and 1 Gbps bandwidth on each side. This experiment opens the door for comparison between different operational modes using the same tool.

\begin{figure}[h!]
	\centering
	\begin{subfigure}[b]{0.48\textwidth}
		\centering
		\resizebox{0.75\linewidth}{!}{%
			\begin{tikzpicture}[node distance=1cm, baseline=(sw.center)]
				\node[simnode] (src) {src};
				\node[simnode, right=of src] (sw)  {sw};
				\node[simnode, right=of sw]  (dst) {dst};
				\draw[simlink] (src) -- node[linklabel, above] {5\,ms} (sw);
				\draw[simlink] (sw)  -- node[linklabel, above] {5\,ms} (dst);
			\end{tikzpicture}%
		}
		\caption{\small Pure Simulation}
		\label{fig:topo-sim}
	\end{subfigure}
	\hfill
	\begin{subfigure}[b]{0.48\textwidth}
		\centering
		\resizebox{0.75\linewidth}{!}{%
			\begin{tikzpicture}[node distance=1cm, baseline=(sw.center)]
				\node[emunode] (src) {src};
				\node[emunode, right=of src] (sw)  {sw};
				\node[emunode, right=of sw]  (dst) {dst};
				\draw[emulink] (src) -- node[linklabel, above] {5\,ms} (sw);
				\draw[emulink] (sw)  -- node[linklabel, above] {5\,ms} (dst);
			\end{tikzpicture}%
		}
		\caption{\small Pure Emulation}
		\label{fig:topo-emu}
	\end{subfigure}
	
	\vspace{0.3em}
	
	\begin{subfigure}[b]{0.48\textwidth}
		\centering
		\resizebox{0.75\linewidth}{!}{%
			\begin{tikzpicture}[node distance=1cm, baseline=(sw.center)]
				\node[emunode] (src) {src};
				\node[simnode, right=of src] (sw)  {sw};
				\node[emunode, right=of sw]  (dst) {dst};
				\draw[bridgelink] (src) -- node[linklabel, above] {5\,ms} (sw);
				\draw[bridgelink] (sw)  -- node[linklabel, above] {5\,ms} (dst);
			\end{tikzpicture}%
		}
		\caption{\small Hybrid}
		\label{fig:topo-hybrid}
	\end{subfigure}
	\hfill
	\begin{subfigure}[b]{0.48\textwidth}
		\centering
		\resizebox{0.8\linewidth}{!}{%
			\begin{tikzpicture}[baseline=(ls.center)]
				\node[simnode, minimum width=0.6cm, minimum height=0.35cm] (ls) {\phantom{x}};
				\node[font=\scriptsize, right=0.2cm of ls] {Simulated node};
				\node[emunode, minimum width=0.6cm, minimum height=0.35cm,
				below=0.4cm of ls] (le) {\phantom{x}};
				\node[font=\scriptsize, right=0.2cm of le] {Emulated node};
				\coordinate (lx) at ([xshift=2.6cm]ls.west);
				\draw[simlink]    (lx)                -- ++(0.8cm,0)
				node[right, font=\scriptsize] {Simulated Link};
				\draw[bridgelink] ([yshift=-0.4cm]lx) -- ++(0.8cm,0)
				node[right, font=\scriptsize] {Bridge Link};
				\draw[emulink]    ([yshift=-0.8cm]lx) -- ++(0.8cm,0)
				node[right, font=\scriptsize] {Emulated Link};
			\end{tikzpicture}%
		}
	\end{subfigure}
	\caption{Topology used in Exp1}
	\label{Expl1-Topology}
\end{figure}

\tikzset{
	simnode/.style={
		draw, rounded corners=3pt, fill=blue!12,
		minimum width=1.2cm, minimum height=0.6cm, inner sep=2pt,
		font=\small\bfseries
	},
	emunode/.style={
		draw, fill=orange!20,
		minimum width=1.2cm, minimum height=0.6cm, inner sep=2pt,
		font=\small\bfseries
	},
	simlink/.style={thick, blue!50, <->, >=Stealth},
	bridgelink/.style={thick, violet!60, dashed, <->, >=Stealth},
	emulink/.style={thick, teal!70, <->, >=Stealth},
	linklabel/.style={font=\small, fill=white, inner sep=1.5pt},
}

\begin{figure}[h!]
\centering
\includegraphics[scale=0.3]{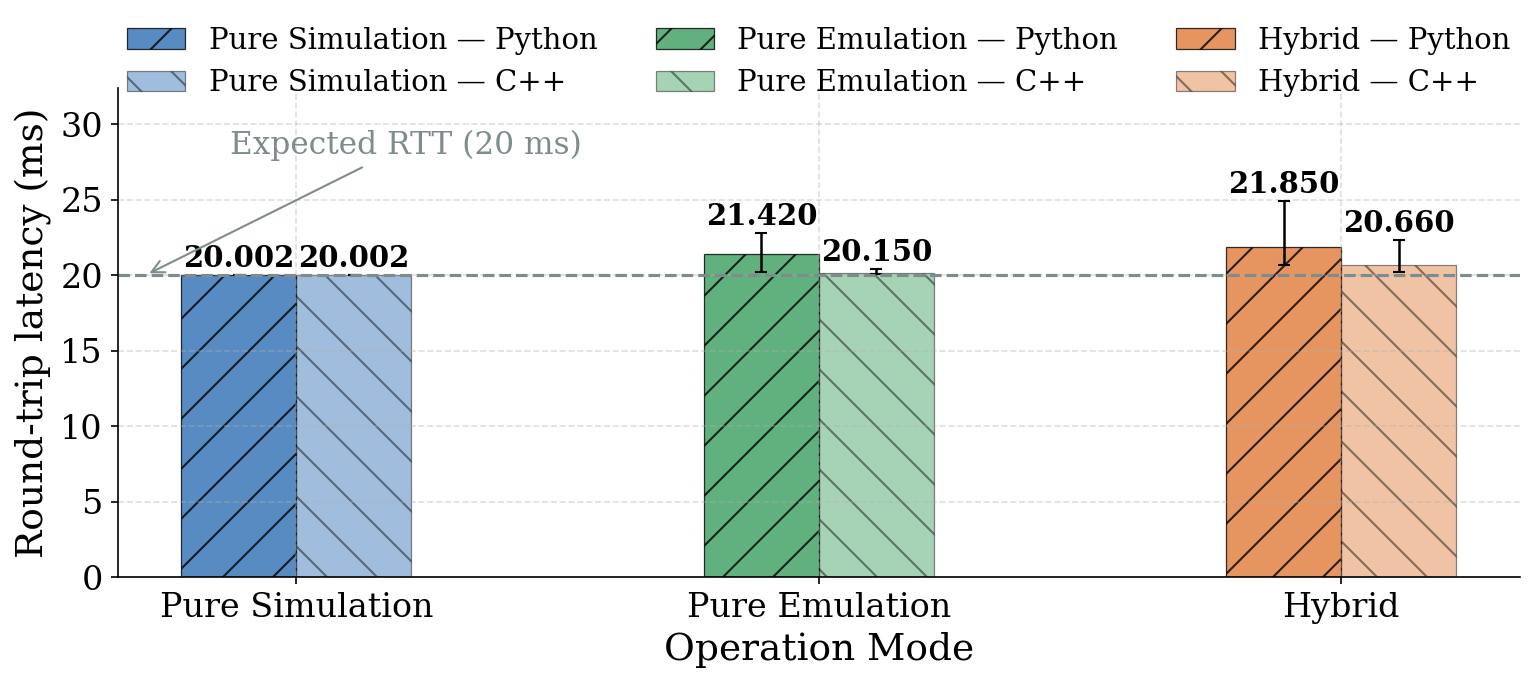}
\caption{Results of Exp1}
\label{Exp1-Latency}
\end{figure}

We also use the AI coding agent to create a C++ version of \ProjectName. Initially, we measure the Round-Trip Time (RTT) delay (see Fig. \ref{Exp1-Latency}). The simulation offers the most accurate results compared with the hybrid and emulation modes. In pure simulation mode, the measured RTT exceeds the expected result by 2µs. The results show that C++ gives lower RTT overhead compared to the Python version for hybrid and emulation modes.

\subsection{Exp2: Testing the pure simulation mode}
\begin{figure}[h!]
\centering
\includegraphics[scale=0.25]{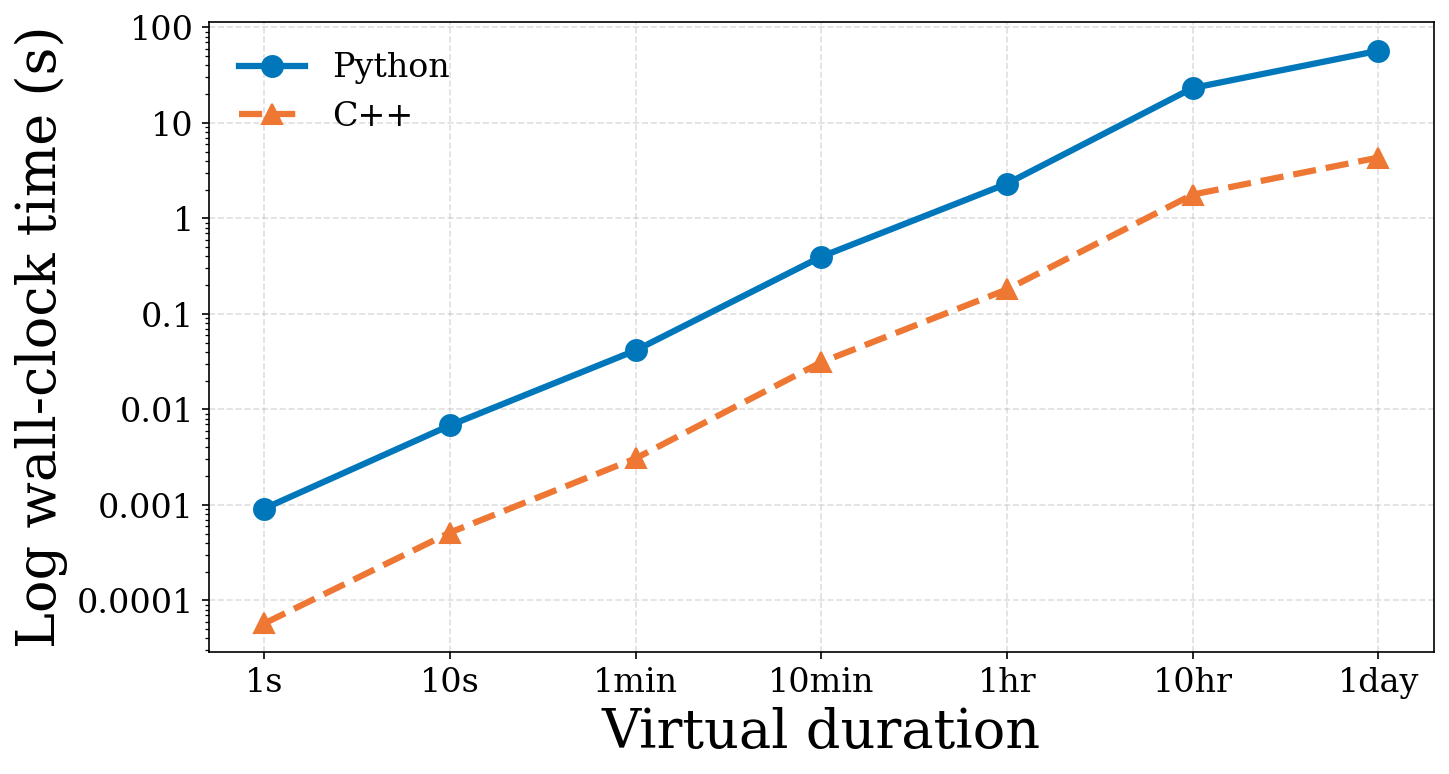}
\caption{Results of Exp2}
\label{Exp2}
\end{figure}

We measure wall-clock time in pure simulation mode across varying virtual durations. We use the same topology as Fig. \ref{Expl1-Topology}(a). A traffic source sends UDP packets at 100 packets per second (pps) (i.e., one packet every 10ms) through a simulated switch to a destination. The virtual time starts from 1 second up to 1 full day. The results show that a full day of network traffic is completed in ~100 seconds (see Fig. \ref{Exp2}) using the Python version. Furthermore, the C++ version shows that 1 day of virtual experiment can be completed in ~4.3 seconds. 
To measure the throughput of \ProjectName, we use the same topology but with different link parameters (100 Gbps, 1 ms delay). The MAC addresses are pre-populated in the switch forwarding table to avoid flooding. The packet rate increases from 100 to 1,000,000 pps. The results for throughput measurements are given in Table \ref{BenchMarktable}. The C++ version shows better throughput compared to the Python version.

\begin{table}[h!]
\centering
\caption{Throughput benchmark}
\label{BenchMarktable}
\setlength{\tabcolsep}{6pt}
\begin{tabular}{r r r r r}
\toprule
\multirow{2}{*}{\textbf{Rate (pps)}}
  & \multicolumn{2}{c}{\textbf{Python}}
  & \multicolumn{2}{c}{\textbf{C++}} \\
\cmidrule(lr){2-3}\cmidrule(lr){4-5}
  & Time (s) & Events/s (K)
  & Time (s) & Events/s (M) \\
\midrule
100           & 0.007  & 299   & 0.001  & 2.73 \\
1K            & 0.066  & 303   & 0.007  & 2.81 \\
10K           & 0.718  & 279   & 0.077  & 2.59 \\
100K          & 8.126  & 246   & 0.852  & 2.35 \\
1M            & 85.690 & 233   & 9.967  & 2.01 \\
\bottomrule
\end{tabular}
\end{table}

\begin{figure}[b!]
	\centering
	\includegraphics[scale=0.27]{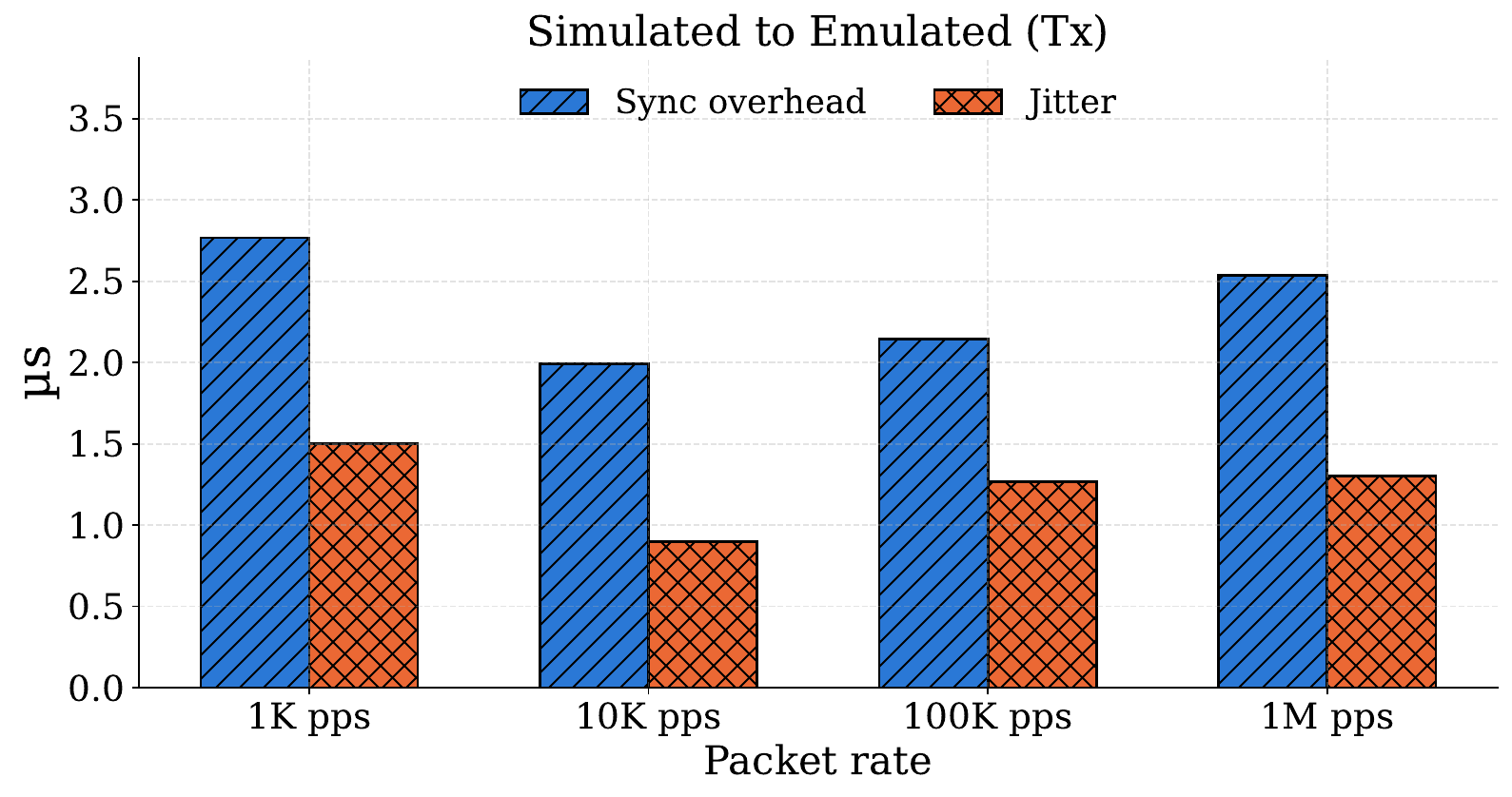}
	\caption{Synchronization overhead and jitter for\\ the simulated-to-emulated direction}
	\label{Sim-Emu}
\end{figure}

\subsection{Exp3: Testing synchronization overhead}
\begin{figure}[h!]
	\centering
	\includegraphics[scale=0.22]{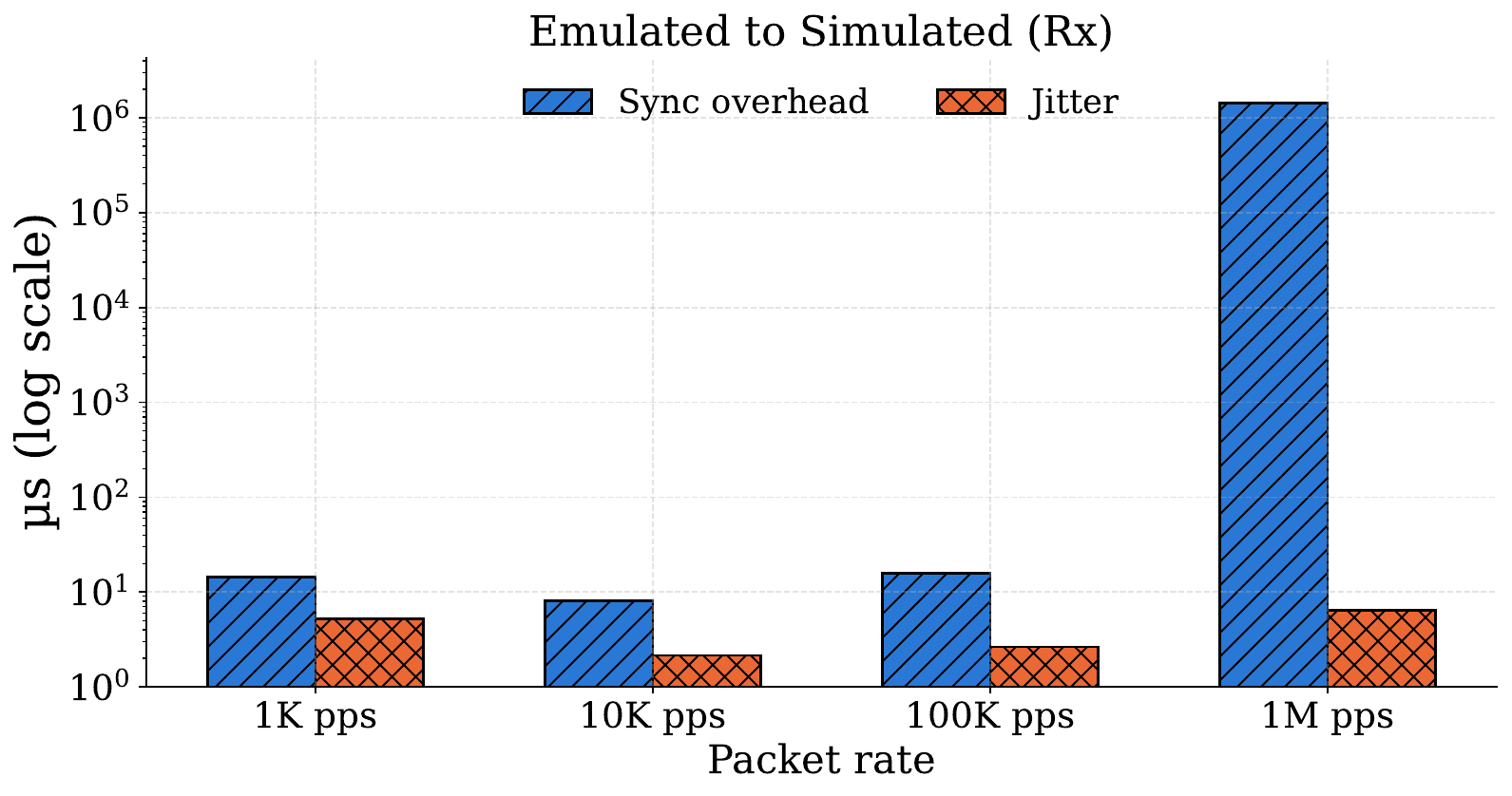}
	\caption{Synchronization overhead and jitter for\\ the emulated- to-simulated direction}
	\label{Emu-Sim}
\end{figure}

We measure the synchronization overhead and jitter for both parts using our C++ version (see Fig. \ref{Sim-Emu} and Fig. \ref{Emu-Sim}). We use the topology for Fig(c). As shown in Fig. \ref{Sim-Emu}, for the simulated-to-emulated direction, we observe that the synchronization overhead is below 3$\mu$s. However, for the emulated-to-simulated direction, as shown in Fig. \ref{Emu-Sim}, we note that the synchronization overhead increases significantly for high packet rates (i.e., 1M packets per second). Jitter values remain low for both cases.

\begin{table*}[h!]
	\centering
	\caption{Results for the generation phase}
	\label{tab:results-gen}
	\begin{tabular}{cl ccc}
		\toprule
		\textbf{Experiment} & \textbf{Description} & \textbf{gpt-5.5} & \textbf{sonnet-5} & \textbf{gemini-3.6-flash} \\
		\midrule
		E1 & rate-limit forwarding under overload            & \texttt{stage3\_fail} & pass                   & \texttt{stage3\_fail} \\
		E2 & token-bucket burst shaping                       & pass                   & pass                   & pass                   \\
		E3 & strict-priority queuing (two traffic classes)     & pass                   & pass                   & pass                   \\
		E4 & L2 MAC-learning switch chain                     & pass                   & pass                   & pass                   \\
		E5 & distance-vector shortest path (4-node mesh)       & pass                   & \texttt{stage3\_fail} & \texttt{stage3\_fail} \\
		E6 & fair queuing across 3 flows                      & pass                   & pass                   & pass                   \\
		E7 & primary/backup link failover                     & pass                   & \texttt{stage3\_fail} & pass                   \\
		E8 & queueing latency under Poisson traffic        & pass                   & pass                   & pass                   \\
		E9 & distance-vector re-convergence after link failure & pass                  & \texttt{stage3\_fail} & \texttt{stage3\_fail} \\
		\midrule
		\textbf{Pass Rate} & & \textbf{8/9} & \textbf{6/9} & \textbf{6/9} \\
		\bottomrule
	\end{tabular}
\end{table*}

\begin{figure}[b!]
	\centering
	\includegraphics[scale=0.665]{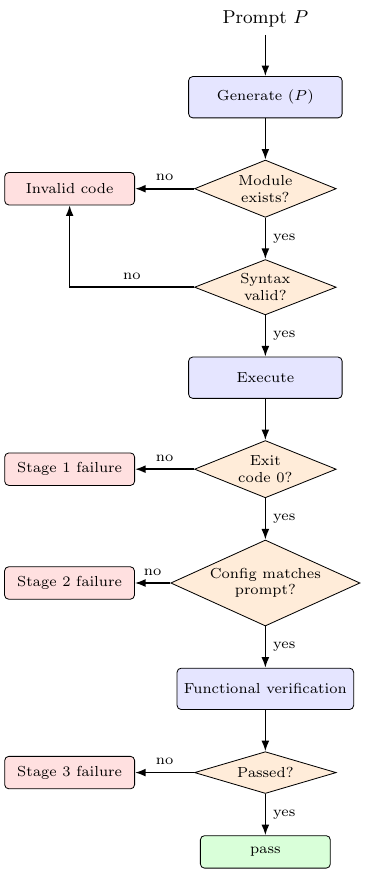}
	\caption{Three stages for the verification phase in AgenticNet}
	\label{AgentVerfError}
\end{figure}

\section{Verifying agent-generated experiments}
This part focuses on the following question: \textit{how can we verify that the agent-generated code correctly implements the experiment specified in a given user's prompt?} Note that we assume that the user's prompt is unambiguous, thus the errors mainly occur at the code level. 
We consider a two-stage approach, in which we separate the code generation phase from the verification phase. We include two agents: one focuses on the generation phase, and another focuses on the verification phase. We consider that the verification agent has knowledge on the user prompt. Initially, the generation agent has to generate new code based on a given user prompt. The agent supports multiple LLM models via a unified API interface. The LLM is provided with \ProjectName{} API definitions and coding examples. After the code generation step, the verification agent performs the following three stages (see Fig. \ref{AgentVerfError}).

\begin{itemize}
	\item \textbf{Stage 1 - syntax and executability:} we ensure that the code uses valid simulator module(s) and is syntactically valid. Then, we ensure that the code is executable. If stage 1 fails, then the code is regenerated.
	\item \textbf{Stage 2 - configuration verification:} we ensure that the configuration in the generated code matches the configuration in the user's prompt. 
	\item \textbf{Stage 3 - functional verification:} for each generated code, a functional validation step is needed to ensure that the experiment is consistent with expected network behavior.
\end{itemize}

\begin{table}[b!]
	\centering
	\caption{Results for the verification phase}
	\label{tab:verifier-results}
	\begin{tabular}{lcc}
		\toprule
		\textbf{Verifier} & \textbf{Iteration Budget} & \textbf{Match Rate} \\[3pt]
		\midrule
		\texttt{gemini-3.6-flash} & 12 & 7/9 \\
		\texttt{gpt-5.5}          & 12 & 8/9 \\
		\texttt{gemini-3.6-flash} & 20 & 9/9 \\
		\texttt{gpt-5.5}          & 20 & 9/9 \\
		\bottomrule
	\end{tabular}
\end{table}

\textbf{Testing the agent-assisted generation phase:}
For simplicity, we use generative AI (Claude Code) to generate 9 simulation-related tasks, which are fed to a single-shot LLM. Our goal is to test the ability of different LLM-based agents to generate experiment code that passes our detection stages. We check whether the code generated using the LLM-based generation approach can pass our user-assisted verification according to the detection stages. The results for the generation phase are shown in Table \ref{tab:results-gen}. The results show that gpt-5.5 shows the best performance in terms of pass rate (8/9). 
Both sonnet-5 and gemini-3.6-flash have lower performance in terms of pass rate (6/9) compared to gpt-5.5. The results show that the generators based on sonnet-5 and gemini-3.6-flash mainly fail during the functional verification stage.

\textbf{Testing the agent-assisted verification phase:}
We also test the agent-assisted verification approach. We check whether the agent-based approach matches the user-assisted verification. We use an iterative agentic verifier, which has access to the simulator tool. We choose codes from the previous generation phase generated using sonnet-5. Those generated codes are examined using our agent-assisted verification approach. We check whether the agent-assisted approach matches the user-assisted verification. The results for the verification stage are shown in Table \ref{tab:verifier-results}. We observe that gpt-5.5 better match rate for a short iteration budget compared to gemini-3.6-flash. Both verifiers required a higher iteration budget to reach better match rate. Overall, the results show that the performance of the generation and verification stages depends mainly on the large language model used by the agent.

\section{Conclusion}
In this paper, we explore the usage of AI coding agents for creating hybrid network experiments. The tool supports three modes: pure simulation, pure emulation, and hybrid mode. \ProjectName{} provides a more flexible approach to creating experiments for cases that may require a combination of simulation and emulation. We explore Python and C++ versions. The results show that C++ offers better performance than the Python version. We also introduce an agent-assisted approach to verify the generated code specified through a user prompt.

\bibliographystyle{ACM-Reference-Format}
\bibliography{sample-base}

\end{document}